\pdfoutput=1

\documentclass[11pt,twoside,a4paper,cmspaper,final,collab]{cms-tdr}
\def\svnVersion{490815P}\def\cmsCernNoTag{CERN-EP-2018-200}\def\cmsCernDate{\today}\def\cmsMessage{Published in Physical Review Letters as \href{http://dx.doi.org/10.1103/PhysRevLett.122.081804}{\doi{10.1103/PhysRevLett.122.081804.}}}

\begin{document}\cmsNoteHeader{EXO-17-019}

\hyphenation{had-ron-i-za-tion}
\hyphenation{cal-or-i-me-ter}
\hyphenation{de-vices}
\RCS$HeadURL: svn+ssh://svn.cern.ch/reps/tdr2/papers/EXO-17-019/trunk/EXO-17-019.tex $
\RCS$Id: EXO-17-019.tex 490702 2019-03-01 21:06:41Z glandsbe $
\providecommand{\cmsTable}[1]{\resizebox{\textwidth}{!}{#1}}
\newlength\cmsFigWidth
\newlength\cmsTabSkip\setlength{\cmsTabSkip}{1ex}
\ifthenelse{\boolean{cms@external}}{\setlength\cmsFigWidth{0.85\columnwidth}}{\setlength\cmsFigWidth{0.4\textwidth}}
\ifthenelse{\boolean{cms@external}}{\providecommand{\cmsLeft}{top\xspace}}{\providecommand{\cmsLeft}{left\xspace}}
\ifthenelse{\boolean{cms@external}}{\providecommand{\cmsRight}{bottom\xspace}}{\providecommand{\cmsRight}{right\xspace}}
\ifthenelse{\boolean{cms@external}}{\providecommand{\CL}{C.L.\xspace}}{\providecommand{\CL}{CL\xspace}}
\newcommand{\MJ}{\ensuremath{m_\mathrm{J}^\text{SD}}\xspace}
\newcommand{\Jg}{\ensuremath{\mathrm{J}\PGg}\xspace}
\newcommand{\Hg}{\ensuremath{\PH\PGg}\xspace}
\cmsNoteHeader{EXO-17-019}
\title{Search for narrow \texorpdfstring{$\PH\PGg$}{H gamma} resonances in proton-proton collisions at \texorpdfstring{$\sqrt{s} = 13\TeV$}{sqrt(s) = 13 TeV}}

\date{\today}

\abstract{
A search for heavy, narrow resonances decaying to a Higgs boson and a photon (\Hg) has been performed in proton-proton collision data at a center-of-mass energy of 13 TeV, corresponding to an integrated luminosity of 35.9\fbinv collected with the CMS detector at the LHC in 2016. Events containing a photon and a Lorentz-boosted hadronically decaying Higgs boson reconstructed as a single, large-radius jet are considered, and the $\PGg$+jet invariant mass spectrum is analyzed for the presence of narrow resonances. To increase the sensitivity of the search, events are categorized depending on whether or not the large-radius jet can be identified as a result of the merging of two jets originating from $\PQb$ quarks. Results in both categories are found to agree with the predictions of the standard model. Upper limits on the production rate of \Hg resonances are set as a function of their mass in the range of 720--3250\GeV, representing the most stringent constraints to date.}

\hypersetup{%
pdfauthor={CMS Collaboration},%
pdftitle={Search for narrow H gamma resonances in proton-proton collisions at sqrt(s) = 13 TeV},%
pdfsubject={CMS},%
pdfkeywords={CMS, physics, resonances, EXO}}

\maketitle

The CERN LHC has enabled the investigation of numerous theories beyond the standard model (SM), among which are those that predict the production of new heavy bosons that decay into a pair of SM bosons. These resonances are found in a large variety of models, from theories with an extended Higgs sector, such as the two Higgs doublet models~\cite{2HDM}, to models with low-scale gravity, \eg, the Randall--Sundrum model~\cite{RS1,RS2}, and models with strong dynamics~\cite{TC,EL}. A number of searches for resonances decaying into the $\PW\PW$, $\PW\PZ$, $\PZ\PZ$, $\PW\PGg$, or $\PZ\PGg$ channels have been carried out at 7, 8, and 13\TeV center-of-mass energies~\cite{diboson0,diboson00,diboson1,diboson2,diboson3,diboson4,diboson4a,diboson5,diboson5a,diboson6,diboson7,diboson8,diboson9,diboson10,diboson11,diboson12,diboson13,diboson14,diboson15,diboson16} by the ATLAS and CMS Collaborations, and stringent limits on their production have been set. The discovery of a Higgs boson ($\PH$)~\cite{Higgs1,Higgs2,Higgs3} allowed for an extension of these searches to include the V$\PH$ channels, where V stands for a $\PW$ or $\PZ$ boson. Negative results of these searches at 8 and 13\TeV have been reported by CMS~\cite{VH-1,VH1,VH2,VH4,diboson7,VH5} and ATLAS~\cite{VH0,VH3,VH6,VH7}.

In this Letter, we extend the above program by reporting on a search for narrow resonances decaying into a Higgs boson and a photon ($\PH\PGg$). The existence of such resonances has been recently predicted~\cite{Dobrescu:2017sue}, using a $\PZpr \to \Hg$ decay at one-loop level as an example, which we chose as the signal benchmark in the present analysis. Here, $\PZpr$ is a $U(1)$ spin-1 boson, similar to the SM $\PZ$ boson. For the decay of a sufficiently massive $\PZpr$ boson via this channel, the Higgs boson would be produced with a significant Lorentz boost. For the dominant Higgs boson decay mode into a pair of $\PQb$ quarks, recently established by ATLAS and CMS~\cite{Aaboud:2017xsd,Sirunyan:2017elk}, the decay products will be reconstructed as a single large-radius jet with a characteristic substructure that can be exploited to distinguish those jets from jets originating from gluons or light-flavor quarks. The background to this process is dominated by SM $\PGg$+jet production, and the application of "\PQb tagging" techniques to identify jets originating from $\PQb$ quarks (``$\PQb$ quark jets" or ``$\PQb$ jets") can reduce the backgrounds by nearly two orders of magnitude. The remaining backgrounds come from nonresonant $\bbbar\PGg$ production, from $\qqbar\PGg$ production with the light-flavor quarks incorrectly identified as $\PQb$ quarks, and from multijet production with light-flavor jets mistagged as $\PQb$ quark jets in addition to a jet misreconstructed as a photon. Very recently, when the work presented here was nearing completion, ATLAS reported on a first search in this channel~\cite{Aaboud:2018fgi}.

The analysis described in this Letter is performed with data corresponding to an integrated luminosity of 35.9\fbinv recorded with the CMS detector at the LHC in proton-proton (\Pp\Pp) collisions at a center-of-mass energy of 13\TeV in 2016. We focus on $\PZpr$ masses above $\approx$700\GeV where the Higgs boson is produced with a sufficient boost to be efficiently reconstructed as a single, large-radius jet. The analysis uses control samples in data to predict the shape of the background and to optimize the selection. The background is obtained via a fit to data with a functional form determined using these control samples and simulation, and a signal is searched for as a narrow resonance on top of a smoothly falling background. To increase the sensitivity of the search, events are categorized depending on whether or not the large-radius jet can be identified as a result of the merging of two jets originating from $\PQb$ quarks. This categorization uses an advanced $\PQb$ tagging technique developed specifically for searches involving Lorentz-boosted Higgs bosons decaying into $\bbbar$ pairs~\cite{Sirunyan:2017ezt}.

The central feature of the CMS apparatus is a superconducting solenoid of 6\unit{m} internal diameter, providing a magnetic field of 3.8\unit{T}. Within the solenoid volume are a silicon pixel and strip tracker, a lead tungstate crystal electromagnetic calorimeter (ECAL), and a brass and scintillator hadron calorimeter (HCAL), each composed of a barrel and two endcap sections. Forward calorimeters extend the pseudorapidity ($\eta$) coverage provided by the barrel and endcap detectors. Muons are detected in gas-ionization chambers embedded in the steel flux-return yoke outside the solenoid. Events of interest are selected using a two-tiered trigger system~\cite{Khachatryan:2016bia}. The first level, composed of custom hardware processors, uses information from the calorimeters and muon detectors to select events at a rate of around 100\unit{kHz} within a time interval of less than 4\mus. The second level, referred to as the high-level trigger, consists of a farm of processors running a version of the full event reconstruction software optimized for fast processing, and reduces the event rate to less than 1\unit{kHz} before data storage. A more detailed description of the CMS detector, together with a definition of the coordinate system used and the relevant kinematic variables, can be found in Ref.~\cite{Chatrchyan:2008zzk}.

Online, events are selected using a logical OR of two triggers: one requiring a photon candidate with a transverse momentum (\pt) greater than 175\GeV and the other requiring a photon with $\pt > 165\GeV$, which in addition must have the ratio of the energy deposits in the HCAL and ECAL below 0.1. Offline, events are reconstructed using a particle-flow (PF) algorithm~\cite{PFlow}, which aims to reconstruct and identify each individual particle (charged or neutral hadron, photon, electron, or muon) in an event, with an optimized combination of information from the various elements of the CMS detector.

For each event, hadronic jets are clustered from PF candidates using the anti-\kt algorithm~\cite{Cacciari:2008gp,Cacciari:2011ma}, with a distance parameter of 0.8 (``large-radius jet," or J), as implemented in \FASTJET~\cite{Cacciari:2011ma}. The jet momentum is determined as the vectorial sum of all particle momenta in the jet, and is found from simulation to be within 5 to 10\% of the true momentum over the whole \pt spectrum and detector acceptance. Additional $\Pp\Pp$ interactions within the same or nearby bunch crossings (pileup) can contribute additional tracks and calorimetric energy depositions to the jet momentum. To mitigate this effect, the pileup-per-particle identification (PUPPI) algorithm~\cite{puppi} is applied to PF candidates prior to jet clustering.
Jet energy corrections are derived from both simulation and data to make the measured response of jets equal to that of particle-level jets on average~\cite{Khachatryan:2016kdb}. The jet energy resolution amounts typically to
8\% at 100\GeV, and 4\% at 1000\GeV. 

Photons are identified with a multivariate analysis (MVA) classifier~\cite{CMS-EGM-14-001} based on
shower shape variables, isolation sums computed from PF candidates in a cone of radius $\Delta R = 0.3$ in the $\eta$--$\phi$ plane, centered on the photon candidate, and variables that account for the dependencies of the shower shape and isolation variables on pileup. In addition, a conversion-safe electron veto~\cite{CMS-EGM-14-001} is applied. Photon candidates are required to satisfy an MVA working point requirement that corresponds to a typical photon reconstruction and identification efficiency of 90\% in the photon $\pt$ range used in the analysis.

At the preselection level, events are required to contain at least one photon candidate with $\pt ^\PGg> 200\GeV$ and $\abs{\eta^\PGg} < 2.4$, excluding the transition region of $1.44 < \abs{\eta} < 1.57$ between the ECAL barrel and endcap, where the reconstruction is not optimal. At least one large-radius jet with $\pt^\mathrm{J} > 250\GeV$ and $\abs{\eta^\mathrm{J}} < 2.6$ is also required. The photon and jet must be separated by $\Delta R(\PGg,\mathrm{J}) > 1.1$, so as to ensure no overlap between the jet and the photon isolation cone. The triggers are found to be more than 98\% efficient for the signal events passing the preselection. 

In addition, to get a more precise estimate of the Higgs boson jet invariant mass, a jet grooming algorithm known as ``soft drop'" (SD)~\cite{softdrop} is applied. The groomed jet mass ($\MJ$) is computed from the sum of the four-momenta of the remaining jet constituents, which are corrected with the same factor as has already been used in the generic jet reconstruction described above. The typical mass resolution for the Higgs boson jet is 10\%~\cite{Sirunyan:2017isc}.

To take advantage of the dominant Higgs boson decay mode, $\PH \to \PQb\PAQb$, we further classify the events according to the output of the dedicated double b tagging (DBT) algorithm~\cite{Sirunyan:2017ezt}, which attempts to identify a two-prong substructure within a large-radius jet, as well as the likelihood that the two subjets originate from the \PQb quarks. For the latter, several associated variables are used in an MVA tagging algorithm. These variables include the significance of the impact parameters of the tracks relative to the primary vertex, the number and masses of secondary vertices, and a variable that characterizes the system of two secondary vertices, taking into account kinematic properties that help to distinguish $\bbbar$ pairs  produced in massive-particle decays from those originating from gluon splitting. We define two event categories using the ``tight" working point of the DBT algorithm~\cite{Sirunyan:2017ezt}, which corresponds to a 36 (20)\% Higgs boson jet tagging efficiency for a signal mass of 1000 (3000) \GeV and an $\approx$1\% QCD jet mistag probability in events passing the preselection. This working point was found to give the greatest signal sensitivity at lower signal masses, below $\approx$2000\GeV, as it provides powerful background rejection. The events with the leading jet passing the tight working point of the DBT algorithm are assigned to the ``\PQb-tagged" category, while the rest are classified as ``untagged". The untagged category allows the search to maintain optimal sensitivity to high-mass resonances (above $\sim$2000\GeV), for which the background is small and at the same time the DBT efficiency deteriorates compared to that at low masses, as the tracks originating from the secondary vertices become more collimated and harder to resolve.

For each category, two regions in preselected data are defined: the search region (SR), in which the invariant mass of the leading jet is required to be $110 < \MJ < 140\GeV$, centered on the nominal Higgs boson mass, and the sideband (SB) region, which requires $100 < \MJ < 110\GeV$. The SB region is chosen because events within it reproduce the shape of the distributions of dominant backgrounds in the SR for the kinematic variables used in the analysis, as seen in data and also confirmed by the Monte Carlo (MC) simulation. Consequently, the SB region can be used to predict the background shape in the SR, and therefore to optimize the analysis without reliance on background simulation. The lower boundary of the SB region is chosen so as to avoid contamination from $\PW$ and $\PZ$ dijet decays, and the upper boundary is chosen far enough from the nominal Higgs boson mass to minimize possible signal contamination. Since the mass of a large-radius jet is loosely correlated with its \pt, we find that with the preselection kinematic requirements, an adequate description of the background shape in the SR is achieved via the SB region for $\Jg$ masses above 720\GeV. Therefore, we search for narrow \Hg resonances in the 720--3250\GeV range, with the upper range determined by the point beyond which we expect to see no background events in data in either of the two categories.

Simulated $\PZpr$ signal samples are generated based on the benchmark model of Ref.~\cite{Dobrescu:2017sue}, as implemented in the \MGvATNLO 2.3.3~\cite{aMCNLO} MC generator. Signals are generated at leading order (LO) in the mass range of 650--3250\GeV, in steps of 100\GeV (650--850\GeV), 150\GeV (850--2050\GeV), or 400\GeV (2050--3250\GeV). The intrinsic width of the signal is chosen to be negligibly small compared to the experimental resolution. The signal shape is parametrized with the Crystal Ball function~\cite{CrystalBall}, which provides an adequate description in the entire mass range considered. For other values of the mass, the signal shape is smoothly interpolated between the Crystal Ball function parameter values derived for the simulated mass points~\cite{Read}. 

While the background estimates come from data, we use simulated background samples to check that the kinematic variables used in the  optimization are adequately described by the data in the SB region. The following simulated background samples are used: multijets, $\PGg$+jets, $\PW$+jets, and $\PZ/\PGg^*$+jets, all generated at LO with \MGvATNLO. All the background samples are normalized using next-to-leading order cross sections and the integrated luminosity of the data sample.

{\tolerance=800 The NNPDF3.0~\cite{Ball:2014uwa} parton distribution functions (PDFs) are used for all simulated samples. The fragmentation and hadronization are described with \PYTHIA version 8.205~\cite{Sjostrand:2014zea} using the underlying event tune CUETP8M1~\cite{CUETP8M1}.  The CMS detector response is simulated using \GEANTfour~\cite{Geant}. All simulated samples include effects of pileup by superimposing on hard scattering events simulated minimum bias collisions, with their multiplicity matching that observed in data.\par}

Scale factors are applied to simulated samples to remove discrepancies between various efficiencies in simulation compared to those in data. These scale factors range 0.85--0.91~\cite{Sirunyan:2017ezt} for the DBT efficiency, and $\approx$0.99~\cite{CMS-EGM-14-001} for the photon identification efficiency.

Further requirements on the kinematic variables of selected events are applied in order to ensure optimal signal sensitivity. The background estimate for the optimization studies comes from the SB region, normalized to the overall number of events in the SR. The optimization is performed separately for the \PQb-tagged and untagged categories. Both the signal significance and the signal limit optimization were checked and in most of the cases the optimal points are the same. Several kinematic variables were considered, including the $N$-subjettiness variables~\cite{nsubjettiness}, characterizing the substructure of the jet. As a result, the following additional requirements were chosen on top of the preselection for both \PQb-tagged and untagged categories: the leading jet must have $\abs{\eta} < 2.2$, ensuring that the core of the jet is within the tracker coverage, the photon must be found in the ECAL barrel ($\abs{\eta} < 1.44$), and the ratio of the photon \pt to the mass of the \Jg system, $\pt^\PGg/m_{\Jg}$, must exceed 0.35. The last two requirements suppress photons from the $\PGg$+jet background, which tend to be more forward than the signal photons. 

\begin{figure}[htb!]
\centering
\includegraphics[width=0.44\textwidth]{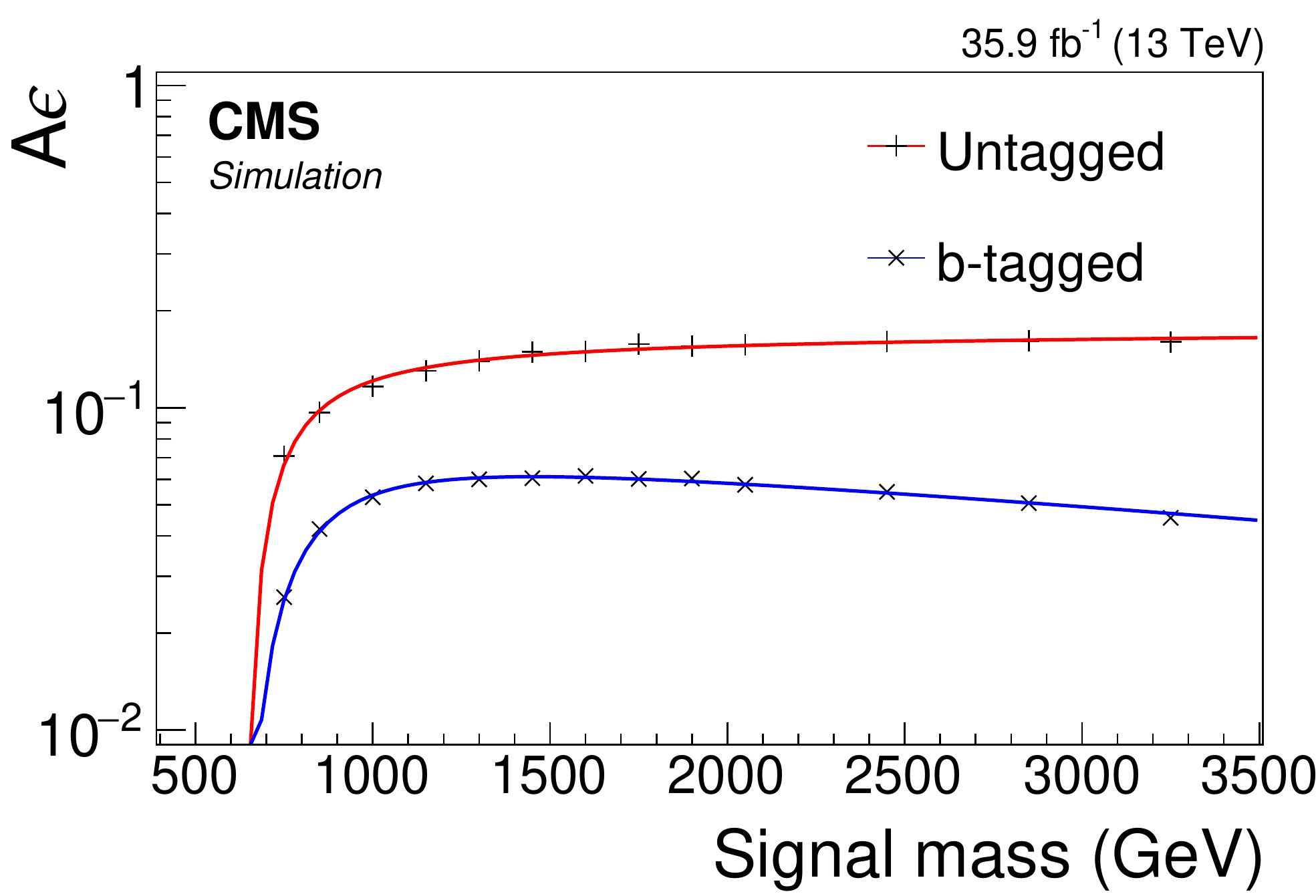}
\caption{Signal acceptance times efficiency ($A\epsilon$) after the final selection, shown for the \PQb-tagged and untagged categories.}
\label{fig:signalFits}
\end{figure}

Shown in Fig.~\ref{fig:signalFits} are the products of signal acceptance and efficiency versus generated signal mass for the two analysis categories, evaluated for each of the simulated signal samples, and a fit function used to interpolate between the generated mass points. In the \PQb-tagged category, the product of the overall acceptance, and the reconstruction, trigger, and full selection efficiency for signal events increases from about 3\% at low signal masses to a peak of 6\% near a signal invariant mass of 1500 \GeV, and decreases thereafter to about 5\% for high signal masses. The observed behavior at high masses is due to the degradation in the DBT efficiency. 
For the untagged category, the corresponding product of acceptance and efficiency increases from about 7\% at low signal masses to around 16\% for high signal masses. The main factors that impact the acceptance are the 57\% SM Higgs boson branching fraction to $\bbbar$ and the fact that about 35\% of the signal large-radius jets fail the $110 < \MJ < 140\GeV$ SR requirement.

After the final selection, the background shape in each category is modeled by fitting a smooth, monotonically falling function to the \Jg invariant mass spectrum in the SR. A variety of functional forms are considered for the background fit, based on the SB region data and on simulated background samples. For every function, a goodness-of-fit (GOF) test known as the method of saturated models~\cite{Cousins} is performed in the SR. The nominal background fit function is then chosen as the one with the best GOF with the minimal number of parameters. The selection of the nominal fit function is performed independently for the \PQb-tagged and untagged categories. In both categories, the following function was found to give the best GOF to the \Jg invariant mass spectrum in the SR: $\rd N/\rd m = p_0(m/\sqrt{s})^{p_1+p_2\log(m/\sqrt{s})}$,  where $p_i$, $i = 0,$ 1, 2 are the free parameters of the fit.

In order to prove that no systematic bias arises because of the choice of the background fit function, a number of tests are performed. An alternative fit function that performed well in the GOF test is used to generate a large number of \Jg invariant mass spectra, with or without signal injection. The spectra are then fit to the sum of the chosen background template and a signal with the mass and normalization allowed to float. The signal significance is extracted from each fit and the distributions of the pull of the signal yield are constructed, where the pull is defined as the difference between the injected and extracted signal normalizations, divided by the statistical uncertainty in the extracted signal normalization from the fit. We observe that the distributions of the pulls are consistent with a Gaussian function with zero mean and a standard deviation of unity, and thus conclude that any systematic bias from the background fitting procedure is negligible compared to the statistical uncertainties in the fit. We therefore use the latter as the only uncertainties associated with the background estimate.

\begin{figure*}[htb!]
\centering
\includegraphics[width=0.45\textwidth]{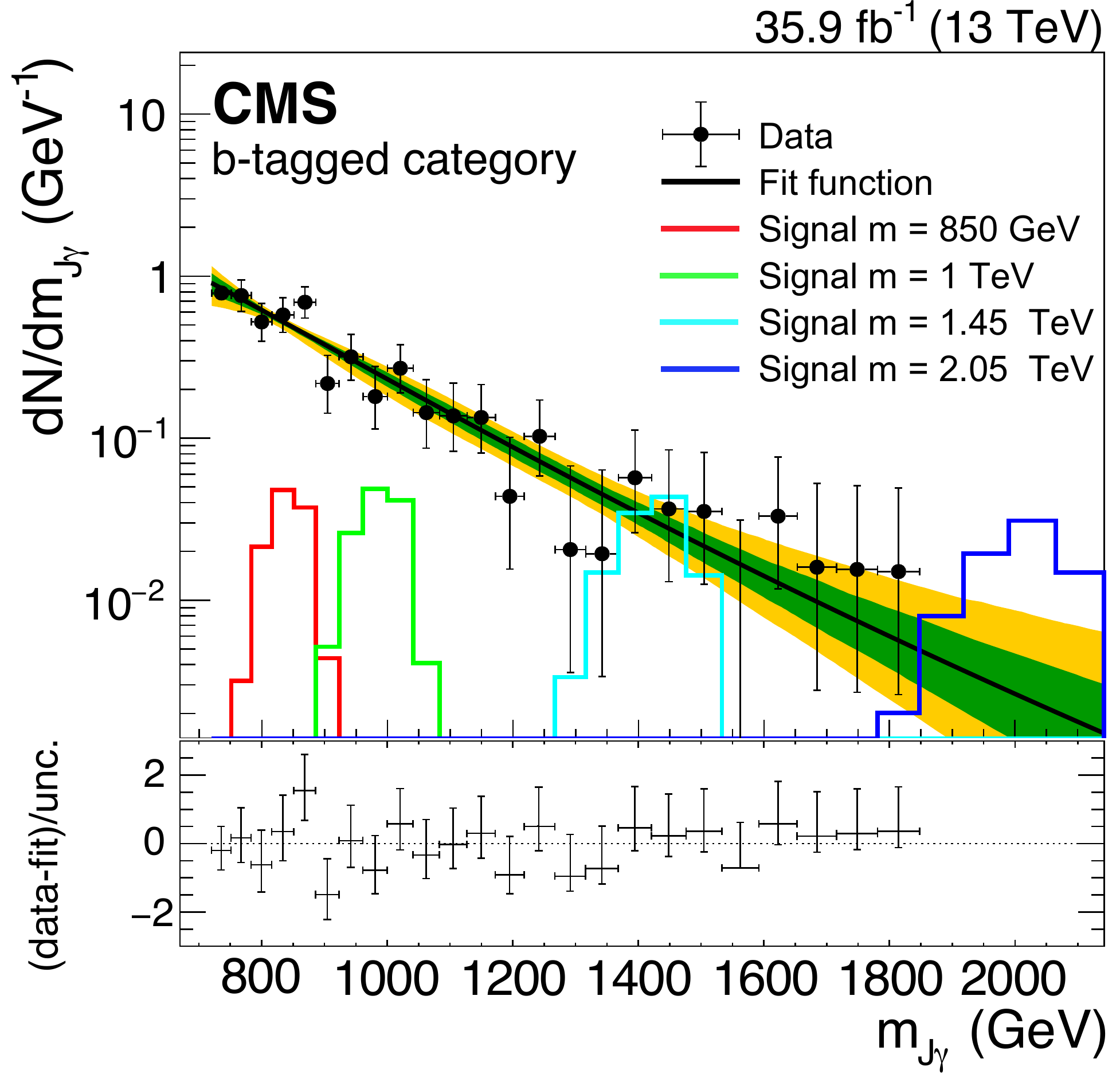}
\includegraphics[width=0.45\textwidth]{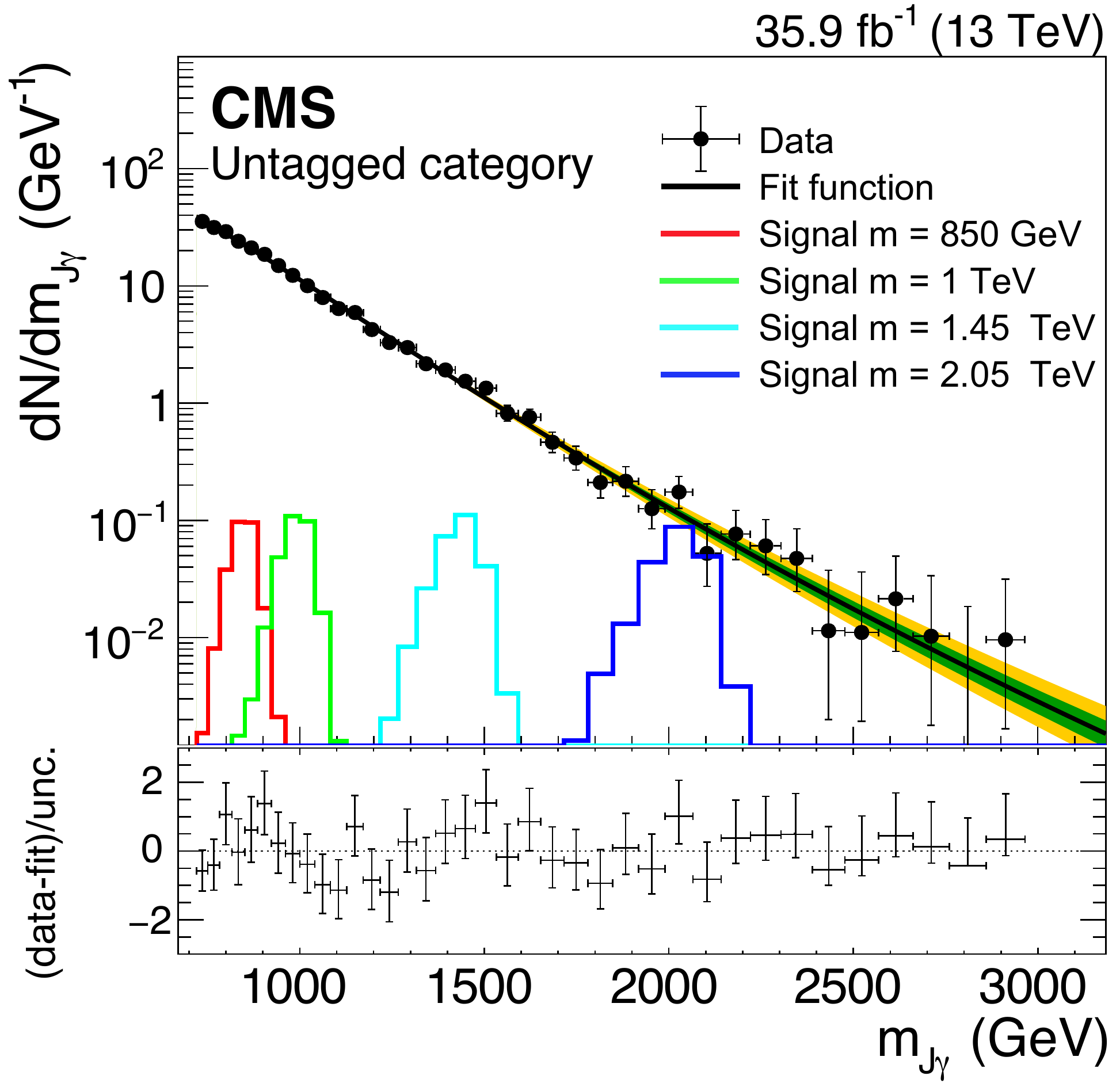}
\caption{The observed \Jg invariant mass spectra in the signal region, shown along with the background fit and a few selected signals, for the \PQb-tagged (left) and untagged (right) categories. Signal samples are plotted with arbitrary normalizations and are shown for illustration purposes. The green and yellow bands correspond to the one and two standard deviation uncertainties in the background-only fit. For bins with a low number of data entries, the error bars correspond to the Garwood confidence intervals~\protect\cite{Garwood}. Shown in the lower panels are the differences between the number of events in data and the nominal background prediction from the fit, divided by the combined statistical uncertainty in the data and the background fit. The error bars correspond to the statistical uncertainty in the data alone.}
\label{fig:ftest}
\end{figure*}

Several systematic uncertainties are taken into account in the signal extraction procedure. These uncertainties stem from effects that may lead to an imperfect estimate of the signal rate and shape, including experimental uncertainties in the integrated luminosity (2.5\%)~\cite{CMS-PAS-LUM-17-001}), jet energy scale and resolution  (2.0\%)~\cite{Chatrchyan:2011ds,Khachatryan:2016kdb}, photon energy scale and resolution (0.1--2.3\%, depending on the \Jg mass~\cite{diboson13}), pileup (1.0\%), groomed jet mass scale (5.0\%), and various identification efficiencies (4.0\%, dominated by the DBT efficiency uncertainty~\cite{Sirunyan:2017ezt}). We also include an uncertainty in signal acceptance due to the PDF choice (2.0\%), based on the PDF4LHC recommendations~\cite{PDF4LHC} using the NNPDF3.0 replicas~\cite{Ball:2014uwa}. Since the correction for the trigger inefficiency in data never exceeds 2\%, the uncertainty due to this correction is always much smaller than the statistical uncertainty of data and therefore has been ignored.

The \Jg invariant mass spectra for both \PQb-tagged and untagged categories, together with the background-only fit, as well as expected signal shapes for several signal masses, are shown in Fig.~\ref{fig:ftest}. The background fit function is shown with the 68 and 95\% confidence level (\CL) uncertainty bands obtained from the fit.  Results in both categories are found to agree with the background-only hypothesis and do not exhibit any significant resonance-like structures. We set upper limits on the production cross section of narrow spin-1 resonances using the modified frequentist \CLs\ criterion~\cite{Junk:1999kv,Read:2002hq,ATL-PHYS-PUB-2011-011}, with a likelihood ratio used as a test statistic, and uncertainties incorporated as nuisance parameters with log-normal priors.

\begin{figure}[htb!]
\centering
\includegraphics[width=0.45\textwidth]{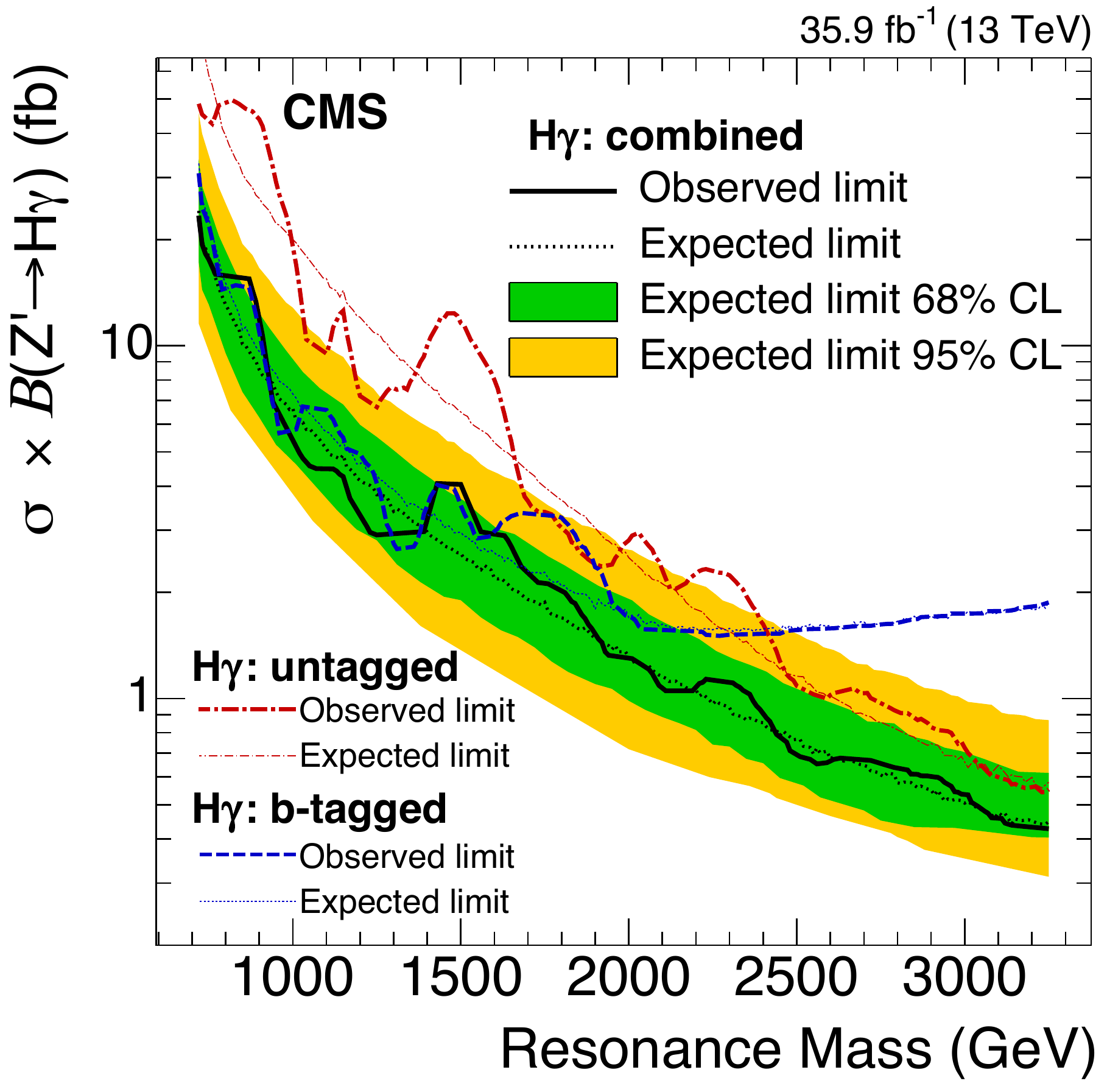}
\caption{Upper limits at 95\% \CL on the product of the signal cross section and the branching fraction to \Hg for the \PQb-tagged and untagged categories, as well as the statistical combination of the two categories.}
\label{fig:limits}
\end{figure}

Shown in Fig.~\ref{fig:limits} are 95\% \CL upper limits on the product of the signal cross section and the branching fraction to \Hg for the \PQb-tagged category, the untagged category, and the statistical combination of the two categories. These limits are the most stringent to date in the entire mass range studied and are the only available limits for masses below 1000\GeV and above 3000\GeV. The significant improvement in the sensitivity compared to the very recent ATLAS limits~\cite{Aaboud:2018fgi} in the 1000--3000\GeV range results from the application of the more efficient DBT algorithm, at low masses, and from the use of the untagged event category, at high masses.

In summary, a search for heavy, narrow resonances decaying to a Higgs boson and a photon (\Hg) has been performed in proton-proton collision data at a center-of-mass energy of 13\TeV, corresponding to an integrated luminosity of 35.9\fbinv collected with the CMS detector at the LHC in 2016. Events in which a photon and a Lorentz-boosted Higgs boson that decays hadronically and is reconstructed as a single, large-radius jet are considered, and the $\PGg$+jet invariant mass spectrum is analyzed for the presence of narrow resonances. To increase the sensitivity of the search, events are categorized depending on whether the large-radius jet can be identified as a result of the merging of two jets originating from $\PQb$ quarks. The backgrounds, dominated by standard model $\PGg$+jet production, are estimated directly from data, without reliance on simulation. Results in both categories are found to agree with the predictions of the standard model. Upper limits on the production cross section of \Hg resonances ranging from 25 to 0.4\unit{fb} are set as a function of the resonance mass in the range of 720--3250\GeV.  These are the most stringent constraints on narrow, spin-1 \Hg resonances to date in the entire mass range, and the first limits available below 1000\GeV and above 3000\GeV.

 \begin{acknowledgments}
We congratulate our colleagues in the CERN accelerator departments for the excellent performance of the LHC and thank the technical and administrative staffs at CERN and at other CMS institutes for their contributions to the success of the CMS effort. In addition, we gratefully acknowledge the computing centers and personnel of the Worldwide LHC Computing Grid for delivering so effectively the computing infrastructure essential to our analyses. Finally, we acknowledge the enduring support for the construction and operation of the LHC and the CMS detector provided by the following funding agencies: BMBWF and FWF (Austria); FNRS and FWO (Belgium); CNPq, CAPES, FAPERJ, FAPERGS, and FAPESP (Brazil); MES (Bulgaria); CERN; CAS, MoST, and NSFC (China); COLCIENCIAS (Colombia); MSES and CSF (Croatia); RPF (Cyprus); SENESCYT (Ecuador); MoER, ERC IUT, and ERDF (Estonia); Academy of Finland, MEC, and HIP (Finland); CEA and CNRS/IN2P3 (France); BMBF, DFG, and HGF (Germany); GSRT (Greece); NKFIA (Hungary); DAE and DST (India); IPM (Iran); SFI (Ireland); INFN (Italy); MSIP and NRF (Republic of Korea); MES (Latvia); LAS (Lithuania); MOE and UM (Malaysia); BUAP, CINVESTAV, CONACYT, LNS, SEP, and UASLP-FAI (Mexico); MOS (Montenegro); MBIE (New Zealand); PAEC (Pakistan); MSHE and NSC (Poland); FCT (Portugal); JINR (Dubna); MON, RosAtom, RAS, RFBR, and NRC KI (Russia); MESTD (Serbia); SEIDI, CPAN, PCTI, and FEDER (Spain); MOSTR (Sri Lanka); Swiss Funding Agencies (Switzerland); MST (Taipei); ThEPCenter, IPST, STAR, and NSTDA (Thailand); TUBITAK and TAEK (Turkey); NASU and SFFR (Ukraine); STFC (United Kingdom); DOE and NSF (USA).
\end{acknowledgments}
\bibliography{auto_generated}

\providecommand{\href}[2]{#2}\begingroup\raggedright\begin{thebibliography}{10}%
\makeatletter
\providecommand{\hrefCMSnoop }[0]{\@secondoftwo}%
\makeatother
\providecommand{\doi}{\texttt{doi:}\begingroup \urlstyle{tt}\Url}

\bibitem{2HDM}
J.~F. Gunion, H.~E. Haber, G.~L. Kane, and S.~Dawson, ``The {Higgs} Hunter's
  Guide'', volume~80 of \textit{ Frontiers in Physics}.
\newblock Westview Press, Boulder, Colorado, 2000.
\newblock
ISBN~073820305X,
%%CITATION = FRPHA,80,1;%%.

\bibitem{RS1}
\hrefCMSnoop {}{L.~Randall and R.~Sundrum, ``A large mass hierarchy from a
  small extra dimension'',} \textit{ Phys. Rev. Lett.} \textbf{ 83} (1999)
  3370,
  \href{http://dx.doi.org/10.1103/PhysRevLett.83.3370}{\doi{10.1103/PhysRevLett.83.3370}},
\href{http://www.arXiv.org/abs/hep-ph/9905221}{\texttt{arXiv:hep-ph/9905221}}.
%%CITATION = HEP-PH/9905221;%%.

\bibitem{RS2}
\hrefCMSnoop {}{L.~Randall and R.~Sundrum, ``An alternative to
  compactification'',} \textit{ Phys. Rev. Lett.} \textbf{ 83} (1999) 4690,
  \href{http://dx.doi.org/10.1103/PhysRevLett.83.4690}{\doi{10.1103/PhysRevLett.83.4690}},
\href{http://www.arXiv.org/abs/hep-th/9906064}{\texttt{arXiv:hep-th/9906064}}.
%%CITATION = HEP-TH/9906064;%%.

\bibitem{TC}
\hrefCMSnoop {}{C.~T. Hill, ``Topcolor assisted technicolor'',} \textit{ Phys.
  Lett. B} \textbf{ 345} (1995) 483,
  \href{http://dx.doi.org/10.1016/0370-2693(94)01660-5}{\doi{10.1016/0370-2693(94)01660-5}},
\href{http://www.arXiv.org/abs/hep-ph/9411426}{\texttt{arXiv:hep-ph/9411426}}.
%%CITATION = HEP-PH/9411426;%%.

\bibitem{EL}
\hrefCMSnoop {}{E.~Eichten and K.~Lane, ``Low-scale technicolor at the
  {Tevatron} and {LHC}'',} \textit{ Phys. Lett. B} \textbf{ 669} (2008) 235,
  \href{http://dx.doi.org/10.1016/j.physletb.2008.09.047}{\doi{10.1016/j.physletb.2008.09.047}},
\href{http://www.arXiv.org/abs/0706.2339}{\texttt{arXiv:0706.2339}}.
%%CITATION = ARXIV:0706.2339;%%.

\bibitem{diboson0}
\hrefCMSnoop {}{{CMS Collaboration}, ``Search for a narrow spin-2 resonance
  decaying to a pair of {Z} vector bosons in the semileptonic final state'',}
  \textit{ Phys. Lett. B} \textbf{ 718} (2013) 1208,
  \href{http://dx.doi.org/10.1016/j.physletb.2012.11.063}{\doi{10.1016/j.physletb.2012.11.063}},
\href{http://www.arXiv.org/abs/1209.3807}{\texttt{arXiv:1209.3807}}.
%%CITATION = ARXIV:1209.3807;%%.

\bibitem{diboson00}
\hrefCMSnoop {}{{CMS Collaboration}, ``Search for exotic resonances decaying
  into {WZ/ZZ} in pp collisions at $\sqrt{s}=7$ {TeV}'',} \textit{ JHEP}
  \textbf{ 02} (2013) 036,
  \href{http://dx.doi.org/10.1007/JHEP02(2013)036}{\doi{10.1007/JHEP02(2013)036}},
\href{http://www.arXiv.org/abs/1211.5779}{\texttt{arXiv:1211.5779}}.
%%CITATION = ARXIV:1211.5779;%%.

\bibitem{diboson1}
\hrefCMSnoop {}{{CMS Collaboration}, ``Search for massive resonances in dijet
  systems containing jets tagged as {W} or {Z} boson decays in pp collisions at
  $\sqrt{s} = 8$ {TeV}'',} \textit{ JHEP} \textbf{ 08} (2014) 173,
  \href{http://dx.doi.org/10.1007/JHEP08(2014)173}{\doi{10.1007/JHEP08(2014)173}},
\href{http://www.arXiv.org/abs/1405.1994}{\texttt{arXiv:1405.1994}}.
%%CITATION = ARXIV:1405.1994;%%.

\bibitem{diboson2}
\hrefCMSnoop {}{{CMS Collaboration}, ``Search for massive resonances decaying
  into pairs of boosted bosons in semi-leptonic final states at $\sqrt{s} = 8$
  {TeV}'',} \textit{ JHEP} \textbf{ 08} (2014) 174,
  \href{http://dx.doi.org/10.1007/JHEP08(2014)174}{\doi{10.1007/JHEP08(2014)174}},
\href{http://www.arXiv.org/abs/1405.3447}{\texttt{arXiv:1405.3447}}.
%%CITATION = ARXIV:1405.3447;%%.

\bibitem{diboson3}
\hrefCMSnoop {}{{ATLAS Collaboration}, ``Combination of searches for {WW},
  {WZ}, and {ZZ} resonances in pp collisions at $\sqrt{s} = 8$ {TeV} with the
  {ATLAS} detector'',} \textit{ Phys. Lett. B} \textbf{ 755} (2016) 285,
  \href{http://dx.doi.org/10.1016/j.physletb.2016.02.015}{\doi{10.1016/j.physletb.2016.02.015}},
\href{http://www.arXiv.org/abs/1512.05099}{\texttt{arXiv:1512.05099}}.
%%CITATION = ARXIV:1512.05099;%%.

\bibitem{diboson4}
\hrefCMSnoop {}{{ATLAS Collaboration}, ``Searches for heavy diboson resonances
  in pp collisions at $\sqrt{s}=13$ {TeV} with the {ATLAS} detector'',}
  \textit{ JHEP} \textbf{ 09} (2016) 173,
  \href{http://dx.doi.org/10.1007/JHEP09(2016)173}{\doi{10.1007/JHEP09(2016)173}},
\href{http://www.arXiv.org/abs/1606.04833}{\texttt{arXiv:1606.04833}}.
%%CITATION = ARXIV:1606.04833;%%.

\bibitem{diboson4a}
\hrefCMSnoop {}{{CMS Collaboration}, ``Search for high-mass {Z$\gamma$}
  resonances in $\mathrm{ e }^{+}\mathrm{ e }^{-}\gamma $ and $
  \mu^{+}\mu^{-}\gamma$ final states in proton-proton collisions at $\sqrt{s}
  =$ 8 and 13 {TeV}'',} \textit{ JHEP} \textbf{ 01} (2017) 076,
  \href{http://dx.doi.org/10.1007/JHEP01(2017)076}{\doi{10.1007/JHEP01(2017)076}},
\href{http://www.arXiv.org/abs/1610.02960}{\texttt{arXiv:1610.02960}}.
%%CITATION = ARXIV:1610.02960;%%.

\bibitem{diboson5}
\hrefCMSnoop {}{{CMS Collaboration}, ``Search for massive resonances decaying
  into {WW}, {WZ} or {ZZ} bosons in proton-proton collisions at $\sqrt{s} = $
  13 {TeV}'',} \textit{ JHEP} \textbf{ 03} (2017) 162,
  \href{http://dx.doi.org/10.1007/JHEP03(2017)162}{\doi{10.1007/JHEP03(2017)162}},
\href{http://www.arXiv.org/abs/1612.09159}{\texttt{arXiv:1612.09159}}.
%%CITATION = ARXIV:1612.09159;%%.

\bibitem{diboson5a}
\hrefCMSnoop {}{{CMS Collaboration}, ``Search for high-mass {$\mathrm{ Z
  }\gamma$} resonances in proton-proton collisions at $\sqrt{s}=$ 8 and 13
  {TeV} using jet substructure techniques'',} \textit{ Phys. Lett. B} \textbf{
  772} (2017) 363,
  \href{http://dx.doi.org/10.1016/j.physletb.2017.06.062}{\doi{10.1016/j.physletb.2017.06.062}},
\href{http://www.arXiv.org/abs/1612.09516}{\texttt{arXiv:1612.09516}}.
%%CITATION = ARXIV:1612.09516;%%.

\bibitem{diboson6}
\hrefCMSnoop {}{{CMS Collaboration}, ``Search for charged {Higgs} bosons
  produced via vector boson fusion and decaying into a pair of {W} and {Z}
  bosons using pp collisions at $\sqrt{s}=13$ {TeV}'',} \textit{ Phys. Rev.
  Lett.} \textbf{ 119} (2017) 141802,
  \href{http://dx.doi.org/10.1103/PhysRevLett.119.141802}{\doi{10.1103/PhysRevLett.119.141802}},
\href{http://www.arXiv.org/abs/1705.02942}{\texttt{arXiv:1705.02942}}.
%%CITATION = ARXIV:1705.02942;%%.

\bibitem{diboson7}
\hrefCMSnoop {}{{CMS Collaboration}, ``Combination of searches for heavy
  resonances decaying to {WW}, {WZ}, {ZZ}, {WH}, and {ZH} boson pairs in
  proton-proton collisions at $\sqrt{s}=8$ and 13 {TeV}'',} \textit{ Phys.
  Lett. B} \textbf{ 774} (2017) 533,
  \href{http://dx.doi.org/10.1016/j.physletb.2017.09.083}{\doi{10.1016/j.physletb.2017.09.083}},
\href{http://www.arXiv.org/abs/1705.09171}{\texttt{arXiv:1705.09171}}.
%%CITATION = ARXIV:1705.09171;%%.

\bibitem{diboson8}
\hrefCMSnoop {}{{ATLAS Collaboration}, ``Search for diboson resonances with
  boson-tagged jets in $pp$ collisions at $\sqrt{s}=13$ {TeV} with the {ATLAS}
  detector'',} \textit{ Phys. Lett. B} \textbf{ 777} (2018) 91,
  \href{http://dx.doi.org/10.1016/j.physletb.2017.12.011}{\doi{10.1016/j.physletb.2017.12.011}},
\href{http://www.arXiv.org/abs/1708.04445}{\texttt{arXiv:1708.04445}}.
%%CITATION = ARXIV:1708.04445;%%.

\bibitem{diboson9}
\hrefCMSnoop {}{{ATLAS Collaboration}, ``Searches for heavy {$ZZ$} and {$ZW$}
  resonances in the $\ell\ell qq$ and $\nu\nu qq$ final states in $pp$
  collisions at $\sqrt{s}=13$ {TeV} with the {ATLAS} detector'',} \textit{
  JHEP} \textbf{ 03} (2018) 009,
  \href{http://dx.doi.org/10.1007/JHEP03(2018)009}{\doi{10.1007/JHEP03(2018)009}},
\href{http://www.arXiv.org/abs/1708.09638}{\texttt{arXiv:1708.09638}}.
%%CITATION = ARXIV:1708.09638;%%.

\bibitem{diboson10}
\hrefCMSnoop {}{{ATLAS Collaboration}, ``Search for {$WW/WZ$} resonance
  production in $\ell \nu qq$ final states in $pp$ collisions at $\sqrt{s} =$
  13 {TeV} with the {ATLAS} detector'',} \textit{ JHEP} \textbf{ 03} (2018)
  042,
  \href{http://dx.doi.org/10.1007/JHEP03(2018)042}{\doi{10.1007/JHEP03(2018)042}},
\href{http://www.arXiv.org/abs/1710.07235}{\texttt{arXiv:1710.07235}}.
%%CITATION = ARXIV:1710.07235;%%.

\bibitem{diboson11}
\hrefCMSnoop {}{{ATLAS Collaboration}, ``Search for heavy resonances decaying
  into {$WW$} in the $e\nu\mu\nu$ final state in $pp$ collisions at
  $\sqrt{s}=13$ {TeV} with the {ATLAS} detector'',} \textit{ Eur. Phys. J. C}
  \textbf{ 78} (2018) 24,
  \href{http://dx.doi.org/10.1140/epjc/s10052-017-5491-4}{\doi{10.1140/epjc/s10052-017-5491-4}},
\href{http://www.arXiv.org/abs/1710.01123}{\texttt{arXiv:1710.01123}}.
%%CITATION = ARXIV:1710.01123;%%.

\bibitem{diboson12}
\hrefCMSnoop {}{{CMS Collaboration}, ``Search for diboson resonances in the
  2$\ell$2$\nu$ final state'',} \textit{ JHEP} \textbf{ 03} (2018) 003,
  \href{http://dx.doi.org/10.1007/JHEP03(2018)003}{\doi{10.1007/JHEP03(2018)003}},
\href{http://www.arXiv.org/abs/1711.04370}{\texttt{arXiv:1711.04370}}.
%%CITATION = ARXIV:1711.04370;%%.

\bibitem{diboson13}
\hrefCMSnoop {}{{CMS Collaboration}, ``{Search for Z$\gamma$ resonances using
  leptonic and hadronic final states in proton-proton collisions at $\sqrt{s}=$
  13 TeV}'',} \textit{ JHEP} \textbf{ 09} (2018) 148,
  \href{http://dx.doi.org/10.1007/JHEP09(2018)148}{\doi{10.1007/JHEP09(2018)148}},
\href{http://www.arXiv.org/abs/1712.03143}{\texttt{arXiv:1712.03143}}.
%%CITATION = ARXIV:1712.03143;%%.

\bibitem{diboson14}
\hrefCMSnoop {}{{CMS Collaboration}, ``Search for a heavy resonance decaying to
  a pair of vector bosons in the lepton plus merged jet final state at
  $\sqrt{s} = 13$ {TeV}'',} \textit{ JHEP} \textbf{ 05} (2018) 088,
  \href{http://dx.doi.org/10.1007/JHEP05(2018)088}{\doi{10.1007/JHEP05(2018)088}},
\href{http://www.arXiv.org/abs/1802.09407}{\texttt{arXiv:1802.09407}}.
%%CITATION = ARXIV:1802.09407;%%.

\bibitem{diboson15}
\hrefCMSnoop {}{{CMS Collaboration}, ``Search for a heavy resonance decaying
  into a {Z} boson and a vector boson in the
  $\nu\overline{\nu}\mathrm{q}\overline{\mathrm{q}}$ final state'',} \textit{
  JHEP} \textbf{ 07} (2018) 075,
  \href{http://dx.doi.org/10.1007/JHEP07(2018)075}{\doi{10.1007/JHEP07(2018)075}},
\href{http://www.arXiv.org/abs/1803.03838}{\texttt{arXiv:1803.03838}}.
%%CITATION = ARXIV:1803.03838;%%.

\bibitem{diboson16}
\hrefCMSnoop {}{{CMS Collaboration}, ``{Search for a heavy resonance decaying
  into a Z boson and a Z or W boson in 2?2q final states at $ \sqrt{s}=13 $
  TeV}'',} \textit{ JHEP} \textbf{ 09} (2018) 101,
  \href{http://dx.doi.org/10.1007/JHEP09(2018)101}{\doi{10.1007/JHEP09(2018)101}},
\href{http://www.arXiv.org/abs/1803.10093}{\texttt{arXiv:1803.10093}}.
%%CITATION = ARXIV:1803.10093;%%.

\bibitem{Higgs1}
\hrefCMSnoop {}{{ATLAS Collaboration}, ``Observation of a new particle in the
  search for the standard model {Higgs} boson with the {ATLAS} detector at the
  {LHC}'',} \textit{ Phys. Lett. B} \textbf{ 716} (2012) 1,
  \href{http://dx.doi.org/10.1016/j.physletb.2012.08.020}{\doi{10.1016/j.physletb.2012.08.020}},
\href{http://www.arXiv.org/abs/1207.7214}{\texttt{arXiv:1207.7214}}.
%%CITATION = ARXIV:1207.7214;%%.

\bibitem{Higgs2}
\hrefCMSnoop {}{{CMS Collaboration}, ``Observation of a new boson at a mass of
  125 {GeV} with the {CMS} experiment at the {LHC}'',} \textit{ Phys. Lett. B}
  \textbf{ 716} (2012) 30,
  \href{http://dx.doi.org/10.1016/j.physletb.2012.08.021}{\doi{10.1016/j.physletb.2012.08.021}},
\href{http://www.arXiv.org/abs/1207.7235}{\texttt{arXiv:1207.7235}}.
%%CITATION = ARXIV:1207.7235;%%.

\bibitem{Higgs3}
\hrefCMSnoop {}{{CMS Collaboration}, ``Observation of a new boson with mass
  near 125 {GeV} in pp collisions at $\sqrt{s} = 7$ and 8 {TeV}'',} \textit{
  JHEP} \textbf{ 06} (2013) 081,
  \href{http://dx.doi.org/10.1007/JHEP06(2013)081}{\doi{10.1007/JHEP06(2013)081}},
\href{http://www.arXiv.org/abs/1303.4571}{\texttt{arXiv:1303.4571}}.
%%CITATION = ARXIV:1303.4571;%%.

\bibitem{VH-1}
\hrefCMSnoop {}{{CMS Collaboration}, ``Search for narrow high-mass resonances
  in proton-proton collisions at $\sqrt{s}$ = 8 {TeV} decaying to a {Z} and a
  {Higgs} boson'',} \textit{ Phys. Lett. B} \textbf{ 748} (2015) 255,
  \href{http://dx.doi.org/10.1016/j.physletb.2015.07.011}{\doi{10.1016/j.physletb.2015.07.011}},
\href{http://www.arXiv.org/abs/1502.04994}{\texttt{arXiv:1502.04994}}.
%%CITATION = ARXIV:1502.04994;%%.

\bibitem{VH1}
\hrefCMSnoop {}{{CMS Collaboration}, ``Search for a massive resonance decaying
  into a {Higgs} boson and a {W} or {Z} boson in hadronic final states in
  proton-proton collisions at $ \sqrt{s}=8$ {TeV}'',} \textit{ JHEP} \textbf{
  02} (2016) 145,
  \href{http://dx.doi.org/10.1007/JHEP02(2016)145}{\doi{10.1007/JHEP02(2016)145}},
\href{http://www.arXiv.org/abs/1506.01443}{\texttt{arXiv:1506.01443}}.
%%CITATION = ARXIV:1506.01443;%%.

\bibitem{VH2}
\hrefCMSnoop {}{{CMS Collaboration}, ``Search for massive {WH} resonances
  decaying into the $\ell \nu \mathrm{b} \overline{\mathrm{b}} $ final state at
  $\sqrt{s}=8$ {TeV}'',} \textit{ Eur. Phys. J. C} \textbf{ 76} (2016) 237,
  \href{http://dx.doi.org/10.1140/epjc/s10052-016-4067-z}{\doi{10.1140/epjc/s10052-016-4067-z}},
\href{http://www.arXiv.org/abs/1601.06431}{\texttt{arXiv:1601.06431}}.
%%CITATION = ARXIV:1601.06431;%%.

\bibitem{VH4}
\hrefCMSnoop {}{{CMS Collaboration}, ``Search for heavy resonances decaying
  into a vector boson and a {Higgs} boson in final states with charged leptons,
  neutrinos, and b quarks'',} \textit{ Phys. Lett. B} \textbf{ 768} (2017) 137,
  \href{http://dx.doi.org/10.1016/j.physletb.2017.02.040}{\doi{10.1016/j.physletb.2017.02.040}},
\href{http://www.arXiv.org/abs/1610.08066}{\texttt{arXiv:1610.08066}}.
%%CITATION = ARXIV:1610.08066;%%.

\bibitem{VH5}
\hrefCMSnoop {}{{CMS Collaboration}, ``Search for heavy resonances that decay
  into a vector boson and a {Higgs} boson in hadronic final states at $\sqrt{s}
  = 13$ {TeV}'',} \textit{ Eur. Phys. J. C} \textbf{ 77} (2017) 636,
  \href{http://dx.doi.org/10.1140/epjc/s10052-017-5192-z}{\doi{10.1140/epjc/s10052-017-5192-z}},
\href{http://www.arXiv.org/abs/1707.01303}{\texttt{arXiv:1707.01303}}.
%%CITATION = ARXIV:1707.01303;%%.

\bibitem{VH0}
\hrefCMSnoop {}{{ATLAS Collaboration}, ``Search for a new resonance decaying to
  a {W} or {Z} boson and a {Higgs} boson in the $\ell \ell / \ell \nu / \nu \nu
  + b \bar{b}$ final states with the {ATLAS} detector'',} \textit{ Eur. Phys.
  J. C} \textbf{ 75} (2015) 263,
  \href{http://dx.doi.org/10.1140/epjc/s10052-015-3474-x}{\doi{10.1140/epjc/s10052-015-3474-x}},
\href{http://www.arXiv.org/abs/1503.08089}{\texttt{arXiv:1503.08089}}.
%%CITATION = ARXIV:1503.08089;%%.

\bibitem{VH3}
\hrefCMSnoop {}{{ATLAS Collaboration}, ``Search for new resonances decaying to
  a {$W$} or {$Z$} boson and a {Higgs} boson in the $\ell^+
  \ell^-\text{b}\bar{\text{b}}$, $\ell \nu\text{b}\bar{\text{b}}$, and
  $\nu\bar{\nu}\text{b}\bar{\text{b}}$ channels with pp collisions at $\sqrt s
  = 13$ {TeV} with the {ATLAS} detector'',} \textit{ Phys. Lett. B} \textbf{
  765} (2017) 32,
  \href{http://dx.doi.org/10.1016/j.physletb.2016.11.045}{\doi{10.1016/j.physletb.2016.11.045}},
\href{http://www.arXiv.org/abs/1607.05621}{\texttt{arXiv:1607.05621}}.
%%CITATION = ARXIV:1607.05621;%%.

\bibitem{VH6}
\hrefCMSnoop {}{{ATLAS Collaboration}, ``Search for heavy resonances decaying
  to a {$W$} or {$Z$} boson and a {Higgs} boson in the
  $q\bar{q}^{(\prime)}b\bar{b}$ final state in $pp$ collisions at $\sqrt{s} =
  13$ {TeV} with the {ATLAS} detector'',} \textit{ Phys. Lett. B} \textbf{ 774}
  (2017) 494,
  \href{http://dx.doi.org/10.1016/j.physletb.2017.09.066}{\doi{10.1016/j.physletb.2017.09.066}},
\href{http://www.arXiv.org/abs/1707.06958}{\texttt{arXiv:1707.06958}}.
%%CITATION = ARXIV:1707.06958;%%.

\bibitem{VH7}
\hrefCMSnoop {}{{ATLAS Collaboration}, ``Search for heavy resonances decaying
  into a {$W$ or $Z$} boson and a {Higgs} boson in final states with leptons
  and $b$-jets in 36 fb$^{-1}$ of $\sqrt{s} = 13$ {TeV} $pp$ collisions with
  the {ATLAS} detector'',} \textit{ JHEP} \textbf{ 03} (2018) 174,
  \href{http://dx.doi.org/10.1007/JHEP03(2018)174}{\doi{10.1007/JHEP03(2018)174}},
\href{http://www.arXiv.org/abs/1712.06518}{\texttt{arXiv:1712.06518}}.
%%CITATION = ARXIV:1712.06518;%%.

\bibitem{Dobrescu:2017sue}
\hrefCMSnoop {}{B.~A. Dobrescu, P.~J. Fox, and J.~Kearney, ``Higgs-photon
  resonances'',} \textit{ Eur. Phys. J. C} \textbf{ 77} (2017) 704,
  \href{http://dx.doi.org/10.1140/epjc/s10052-017-5276-9}{\doi{10.1140/epjc/s10052-017-5276-9}},
\href{http://www.arXiv.org/abs/1705.08433}{\texttt{arXiv:1705.08433}}.
%%CITATION = ARXIV:1705.08433;%%.

\bibitem{Aaboud:2017xsd}
\hrefCMSnoop {}{{ATLAS Collaboration}, ``{Evidence for the $ H\to b\overline{b}
  $ decay with the ATLAS detector}'',} \textit{ JHEP} \textbf{ 12} (2017) 024,
  \href{http://dx.doi.org/10.1007/JHEP12(2017)024}{\doi{10.1007/JHEP12(2017)024}},
\href{http://www.arXiv.org/abs/1708.03299}{\texttt{arXiv:1708.03299}}.
%%CITATION = ARXIV:1708.03299;%%.

\bibitem{Sirunyan:2017elk}
\hrefCMSnoop {}{{CMS Collaboration}, ``{Evidence for the Higgs boson decay to a
  bottom quark-antiquark pair}'',} \textit{ Phys. Lett. B} \textbf{ 780} (2018)
  501,
  \href{http://dx.doi.org/10.1016/j.physletb.2018.02.050}{\doi{10.1016/j.physletb.2018.02.050}},
\href{http://www.arXiv.org/abs/1709.07497}{\texttt{arXiv:1709.07497}}.
%%CITATION = ARXIV:1709.07497;%%.

\bibitem{Aaboud:2018fgi}
\hrefCMSnoop {}{{ATLAS Collaboration}, ``{Search for heavy resonances decaying
  to a photon and a hadronically decaying $Z/W/H$ boson in $pp$ collisions at
  $\sqrt{s}=13$ $\mathrm{TeV}$ with the ATLAS detector}'',} \textit{ Phys. Rev.
  D} \textbf{ 98} (2018) 032015,
  \href{http://dx.doi.org/10.1103/PhysRevD.98.032015}{\doi{10.1103/PhysRevD.98.032015}},
\href{http://www.arXiv.org/abs/1805.01908}{\texttt{arXiv:1805.01908}}.
%%CITATION = ARXIV:1805.01908;%%.

\bibitem{Sirunyan:2017ezt}
\hrefCMSnoop {}{{CMS Collaboration}, ``Identification of heavy-flavour jets
  with the {CMS} detector in pp collisions at 13 {TeV}'',} \textit{ JINST}
  \textbf{ 13} (2018) P05011,
  \href{http://dx.doi.org/10.1088/1748-0221/13/05/P05011}{\doi{10.1088/1748-0221/13/05/P05011}},
\href{http://www.arXiv.org/abs/1712.07158}{\texttt{arXiv:1712.07158}}.
%%CITATION = ARXIV:1712.07158;%%.

\bibitem{Khachatryan:2016bia}
\hrefCMSnoop {}{{CMS Collaboration}, ``The {CMS} trigger system'',} \textit{
  JINST} \textbf{ 12} (2017) P01020,
  \href{http://dx.doi.org/10.1088/1748-0221/12/01/P01020}{\doi{10.1088/1748-0221/12/01/P01020}},
\href{http://www.arXiv.org/abs/1609.02366}{\texttt{arXiv:1609.02366}}.
%%CITATION = ARXIV:1609.02366;%%.

\bibitem{Chatrchyan:2008zzk}
\hrefCMSnoop {}{{CMS Collaboration}, ``The {CMS} experiment at the {CERN}
  {LHC}'',} \textit{ JINST} \textbf{ 3} (2008) S08004,
\href{http://dx.doi.org/10.1088/1748-0221/3/08/S08004}{\doi{10.1088/1748-0221/3/08/S08004}}.
%%CITATION = JINST,3,S08004;%%.

\bibitem{PFlow}
\hrefCMSnoop {}{{CMS Collaboration}, ``Particle-flow reconstruction and global
  event description with the {CMS} detector'',} \textit{ JINST} \textbf{ 12}
  (2017) P10003,
  \href{http://dx.doi.org/10.1088/1748-0221/12/10/P10003}{\doi{10.1088/1748-0221/12/10/P10003}},
\href{http://www.arXiv.org/abs/1706.04965}{\texttt{arXiv:1706.04965}}.
%%CITATION = ARXIV:1706.04965;%%.

\bibitem{Cacciari:2008gp}
\hrefCMSnoop {}{M.~Cacciari, G.~P. Salam, and G.~Soyez, ``The anti-\kt jet
  clustering algorithm'',} \textit{ JHEP} \textbf{ 04} (2008) 063,
  \href{http://dx.doi.org/10.1088/1126-6708/2008/04/063}{\doi{10.1088/1126-6708/2008/04/063}},
  \href{http://www.arXiv.org/abs/0802.1189}{\texttt{arXiv:0802.1189}}.

\bibitem{Cacciari:2011ma}
\hrefCMSnoop {}{M.~Cacciari, G.~P. Salam, and G.~Soyez, ``{FastJet} user
  manual'',} \textit{ Eur. Phys. J. C} \textbf{ 72} (2012) 1896,
  \href{http://dx.doi.org/10.1140/epjc/s10052-012-1896-2}{\doi{10.1140/epjc/s10052-012-1896-2}},
\href{http://www.arXiv.org/abs/1111.6097}{\texttt{arXiv:1111.6097}}.
%%CITATION = ARXIV:1111.6097;%%.

\bibitem{puppi}
\hrefCMSnoop {}{D.~Bertolini, P.~Harris, M.~Low, and N.~Tran, ``Pileup per
  particle identification'',} \textit{ JHEP} \textbf{ 10} (2014) 059,
  \href{http://dx.doi.org/10.1007/JHEP10(2014)059}{\doi{10.1007/JHEP10(2014)059}},
\href{http://www.arXiv.org/abs/1407.6013}{\texttt{arXiv:1407.6013}}.
%%CITATION = ARXIV:1407.6013;%%.

\bibitem{Khachatryan:2016kdb}
\hrefCMSnoop {}{{CMS Collaboration}, ``Jet energy scale and resolution in the
  {CMS} experiment in pp collisions at 8 {TeV}'',} \textit{ JINST} \textbf{ 12}
  (2017) P02014,
  \href{http://dx.doi.org/10.1088/1748-0221/12/02/P02014}{\doi{10.1088/1748-0221/12/02/P02014}},
\href{http://www.arXiv.org/abs/1607.03663}{\texttt{arXiv:1607.03663}}.
%%CITATION = ARXIV:1607.03663;%%.

\bibitem{CMS-EGM-14-001}
\hrefCMSnoop {}{{CMS Collaboration}, ``Performance of photon reconstruction and
  identification with the {CMS} detector in proton-proton collisions at
  $\sqrt{s} = 8$ {TeV}'',} \textit{ JINST} \textbf{ 10} (2015) P08010,
  \href{http://dx.doi.org/10.1088/1748-0221/10/08/P08010}{\doi{10.1088/1748-0221/10/08/P08010}},
\href{http://www.arXiv.org/abs/1502.02702}{\texttt{arXiv:1502.02702}}.
%%CITATION = ARXIV:1502.02702;%%.

\bibitem{softdrop}
\hrefCMSnoop {}{A.~J. Larkoski, S.~Marzani, G.~Soyez, and J.~Thaler, ``Soft
  drop'',} \textit{ JHEP} \textbf{ 05} (2014) 146,
  \href{http://dx.doi.org/10.1007/JHEP05(2014)146}{\doi{10.1007/JHEP05(2014)146}},
\href{http://www.arXiv.org/abs/1402.2657}{\texttt{arXiv:1402.2657}}.
%%CITATION = ARXIV:1402.2657;%%.

\bibitem{Sirunyan:2017isc}
\hrefCMSnoop {}{{CMS Collaboration}, ``{Search for a massive resonance decaying
  to a pair of Higgs bosons in the four b quark final state in proton-proton
  collisions at $\sqrt{s}=$ 13 TeV}'',} \textit{ Phys. Lett. B} \textbf{ 781}
  (2018) 244,
  \href{http://dx.doi.org/10.1016/j.physletb.2018.03.084}{\doi{10.1016/j.physletb.2018.03.084}},
\href{http://www.arXiv.org/abs/1710.04960}{\texttt{arXiv:1710.04960}}.
%%CITATION = ARXIV:1710.04960;%%.

\bibitem{aMCNLO}
J.~Alwall\hrefCMSnoop {}{ {et~al.}, ``The automated computation of tree-level
  and next-to-leading order differential cross sections, and their matching to
  parton shower simulations'',} \textit{ JHEP} \textbf{ 07} (2014) 079,
  \href{http://dx.doi.org/10.1007/JHEP07(2014)079}{\doi{10.1007/JHEP07(2014)079}},
\href{http://www.arXiv.org/abs/1405.0301}{\texttt{arXiv:1405.0301}}.
%%CITATION = ARXIV:1405.0301;%%.

\bibitem{CrystalBall}
\href {http://www.slac.stanford.edu/pubs/slacreports/ slac-r-236.html}{M.~J.
  Oreglia, ``A study of the reactions $\psi^\prime \to \gamma \gamma \psi$''}.
\newblock PhD thesis, Stanford University, 1980.
\newblock {{SLAC} Report {SLAC-R-236}}.

\bibitem{Read}
\hrefCMSnoop {}{A.~L. Read, ``Linear interpolation of histograms'',} \textit{
  Nucl. Instrum. Meth. A} \textbf{ 425} (1999) 357,
\href{http://dx.doi.org/10.1016/S0168-9002(98)01347-3}{\doi{10.1016/S0168-9002(98)01347-3}}.
%%CITATION = NUIMA,A425,357;%%.

\bibitem{Ball:2014uwa}
\hrefCMSnoop {}{{NNPDF} Collaboration, ``Parton distributions for the {LHC Run
  II}'',} \textit{ JHEP} \textbf{ 04} (2015) 040,
  \href{http://dx.doi.org/10.1007/JHEP04(2015)040}{\doi{10.1007/JHEP04(2015)040}},
\href{http://www.arXiv.org/abs/1410.8849}{\texttt{arXiv:1410.8849}}.
%%CITATION = ARXIV:1410.8849;%%.

\bibitem{Sjostrand:2014zea}
T.~Sj{\"o}strand\hrefCMSnoop {}{ {et~al.}, ``An introduction to {PYTHIA
  8.2}'',} \textit{ Comput. Phys. Commun.} \textbf{ 191} (2015) 159,
  \href{http://dx.doi.org/10.1016/j.cpc.2015.01.024}{\doi{10.1016/j.cpc.2015.01.024}},
\href{http://www.arXiv.org/abs/1410.3012}{\texttt{arXiv:1410.3012}}.
%%CITATION = ARXIV:1410.3012;%%.

\bibitem{CUETP8M1}
\hrefCMSnoop {}{{CMS Collaboration}, ``Event generator tunes obtained from
  underlying event and multiparton scattering measurements'',} \textit{ Eur.
  Phys. J. C} \textbf{ 76} (2016) 155,
  \href{http://dx.doi.org/10.1140/epjc/s10052-016-3988-x}{\doi{10.1140/epjc/s10052-016-3988-x}},
\href{http://www.arXiv.org/abs/1512.00815}{\texttt{arXiv:1512.00815}}.
%%CITATION = ARXIV:1512.00815;%%.

\bibitem{Geant}
\hrefCMSnoop {}{{GEANT4} Collaboration, ``{\GEANTfour}---a simulation
  toolkit'',} \textit{ Nucl. Instrum. Meth. A} \textbf{ 506} (2003) 250,
  \href{http://dx.doi.org/10.1016/S0168-9002(03)01368-8}{\doi{10.1016/S0168-9002(03)01368-8}}.

\bibitem{nsubjettiness}
\hrefCMSnoop {}{J.~Thaler and K.~Van~Tilburg, ``Identifying boosted objects
  with {N}-subjettiness'',} \textit{ JHEP} \textbf{ 03} (2011) 015,
  \href{http://dx.doi.org/10.1007/JHEP03(2011)015}{\doi{10.1007/JHEP03(2011)015}},
\href{http://www.arXiv.org/abs/1011.2268}{\texttt{arXiv:1011.2268}}.
%%CITATION = ARXIV:1011.2268;%%.

\bibitem{Cousins}
\hrefCMSnoop {}{S.~Baker and R.~D. Cousins, ``Clarification of the use of
  chi-square and likelihood functions in fits to histograms'',} \textit{ Nucl.
  Instrum. Meth.} \textbf{ 221} (1984) 437,
\href{http://dx.doi.org/10.1016/0167-5087(84)90016-4}{\doi{10.1016/0167-5087(84)90016-4}}.
%%CITATION = NUIMA,221,437;%%.

\bibitem{Garwood}
\hrefCMSnoop {}{F.~Garwood, ``Fiducial limits for the {Poisson}
  distribution'',} \textit{ Biometrika} \textbf{ 28} (1936) 437,
  \href{http://dx.doi.org/10.1093/biomet/28.3-4.437}{\doi{10.1093/biomet/28.3-4.437}}.

\bibitem{CMS-PAS-LUM-17-001}
\href {http://cds.cern.ch/record/2257069}{{CMS Collaboration}, ``{CMS}
  luminosity measurements for the 2016 data-taking period'',} CMS Physics
  Analysis Summary CMS-PAS-LUM-17-001, 2017.

\bibitem{Chatrchyan:2011ds}
\hrefCMSnoop {}{{CMS Collaboration}, ``Determination of jet energy calibration
  and transverse momentum resolution in {CMS}'',} \textit{ JINST} \textbf{ 6}
  (2011) P11002,
  \href{http://dx.doi.org/10.1088/1748-0221/6/11/P11002}{\doi{10.1088/1748-0221/6/11/P11002}},
\href{http://www.arXiv.org/abs/1107.4277}{\texttt{arXiv:1107.4277}}.
%%CITATION = ARXIV:1107.4277;%%.

\bibitem{PDF4LHC}
\hrefCMSnoop {}{J.~Butterworth {et~al.}, ``{PDF4LHC} recommendations for {LHC}
  {Run II}'',} \textit{ J. Phys. G} \textbf{ 43} (2016) 023001,
  \href{http://dx.doi.org/10.1088/0954-3899/43/2/023001}{\doi{10.1088/0954-3899/43/2/023001}},
\href{http://www.arXiv.org/abs/1510.03865}{\texttt{arXiv:1510.03865}}.
%%CITATION = ARXIV:1510.03865;%%.

\bibitem{Junk:1999kv}
\hrefCMSnoop {}{T.~Junk, ``Confidence level computation for combining searches
  with small statistics'',} \textit{ Nucl. Instrum. Meth. A} \textbf{ 434}
  (1999) 435,
  \href{http://dx.doi.org/10.1016/S0168-9002(99)00498-2}{\doi{10.1016/S0168-9002(99)00498-2}},
\href{http://www.arXiv.org/abs/hep-ex/9902006}{\texttt{arXiv:hep-ex/9902006}}.
%%CITATION = HEP-EX/9902006;%%.

\bibitem{Read:2002hq}
\hrefCMSnoop {}{A.~L. Read, ``Presentation of search results: The {\CLs}
  technique'',} \textit{ J. Phys. G} \textbf{ 28} (2002) 2693,
\href{http://dx.doi.org/10.1088/0954-3899/28/10/313}{\doi{10.1088/0954-3899/28/10/313}}.
%%CITATION = JPHGB,G28,2693;%%.

\bibitem{ATL-PHYS-PUB-2011-011}
\href {https://cdsweb.cern.ch/record/1379837}{{ATLAS and CMS Collaborations},
  ``Procedure for the {LHC Higgs} boson search combination in summer 2011'',}
  ATL-PHYS-PUB-2011-011, {CMS NOTE-2011/005}, 2011.

\end{thebibliography}\endgroup
\cleardoublepage \appendix\section{The CMS Collaboration \label{app:collab}}\begin{sloppypar}\hyphenpenalty=5000\widowpenalty=500\clubpenalty=5000\vskip\cmsinstskip
\textbf{Yerevan Physics Institute, Yerevan, Armenia}\\*[0pt]
A.M.~Sirunyan, A.~Tumasyan
\vskip\cmsinstskip
\textbf{Institut f\"{u}r Hochenergiephysik, Wien, Austria}\\*[0pt]
W.~Adam, F.~Ambrogi, E.~Asilar, T.~Bergauer, J.~Brandstetter, M.~Dragicevic, J.~Er\"{o}, A.~Escalante~Del~Valle, M.~Flechl, R.~Fr\"{u}hwirth\cmsAuthorMark{1}, V.M.~Ghete, J.~Hrubec, M.~Jeitler\cmsAuthorMark{1}, N.~Krammer, I.~Kr\"{a}tschmer, D.~Liko, T.~Madlener, I.~Mikulec, N.~Rad, H.~Rohringer, J.~Schieck\cmsAuthorMark{1}, R.~Sch\"{o}fbeck, M.~Spanring, D.~Spitzbart, A.~Taurok, W.~Waltenberger, J.~Wittmann, C.-E.~Wulz\cmsAuthorMark{1}, M.~Zarucki
% ...
\vskip\cmsinstskip
% ...
1:  Also at Vienna University of Technology, Vienna, Austria\\
% ...
\end{sloppypar}
\end{document}